\documentclass[conference]{IEEEtran}
\ifCLASSOPTIONcompsoc
  \usepackage[nocompress]{cite}
\else
  \usepackage{cite}
\fi
\ifCLASSINFOpdf
  \usepackage[pdftex]{graphicx}
\else
\fi

\usepackage{amsmath}

\usepackage[ruled, linesnumbered,noend]{algorithm2e}

\usepackage{array}

\ifCLASSOPTIONcompsoc
  \usepackage[caption=false,font=footnotesize,labelfont=sf,textfont=sf]{subfig}
\else
  \usepackage[caption=false,font=footnotesize]{subfig}
\fi
\usepackage{float}

\usepackage[multiple]{footmisc}
\usepackage{url}
\usepackage{esvect}
\usepackage{textcomp}
\usepackage{diagbox}
\usepackage{tabularx}
\usepackage{dcolumn}
\usepackage{booktabs}
\usepackage[english]{babel}
\usepackage[dvipsnames]{xcolor}
\usepackage[utf8]{inputenc}
\usepackage[T1]{fontenc}
\usepackage{latexsym}
\usepackage{amssymb}
\usepackage{listings}
\usepackage{caption}
\usepackage{subfig}

\usepackage{longtable}
\usepackage{rotating}
\usepackage{array}
\usepackage{multicol}
\usepackage{multirow}

\makeatletter
\newcommand{\removelatexerror}{\let\@latex@error\@gobble}
\makeatother

\makeatletter
\newcommand*{\rom}[1]{\expandafter\@slowromancap\romannumeral #1@}
\makeatother

\lstdefinelanguage{code}{
  morekeywords={let,in,def,aspect,before,after,pointcut,public,privileged,protected,declare,parents, call,target,implements,throw,new,for,class,forAll,exists,Boolean,return,break,executeCaller,true,
  false,pre,if,then,else,endif,String,join,select,from,where,with,create,temporary,table,and,
  or,as,on,case,when,end,union,iterate,intersection,symmetricDifference,self,includes,function, Text,Set,boolean, show,insert, into,div,mod,Integer, by, having, on, not, var, while, continue,switch,execute,abort,delete,eval,this,input, output,process},
  basicstyle=\footnotesize\usefont{T1}{pcr}{m}{n}\selectfont,
  keywordstyle=\footnotesize\usefont{T1}{pcr}{b}{n}\selectfont,
  identifierstyle=\footnotesize\usefont{T1}{pcr}{m}{n}\selectfont,
  commentstyle=,
  stringstyle=\footnotesize\usefont{T1}{pcr}{m}{n}\selectfont,
  numberstyle=\footnotesize,
  tabsize=2,
  frame=lines,
  upquote=true,
  xleftmargin=10pt
}

\usepackage{cite}

\begin{document}

\title{On the Design of Co-operating Blockchains for IoT}

\author{\IEEEauthorblockN{Gokhan Sagirlar}
\IEEEauthorblockA{\textit{IBM Research}\\
Dublin, Ireland \\
gokhan.sagirlar1@ibm.com}
\and
\IEEEauthorblockN{John D. Sheehan}
\IEEEauthorblockA{\textit{IBM Research}\\
Dublin, Ireland \\
john.d.sheehan@ie.ibm.com}
\and
\IEEEauthorblockN{Emanuele Ragnoli}
\IEEEauthorblockA{\textit{IBM Research}\\
Dublin, Ireland \\
eragnoli@ie.ibm.com}
}

\maketitle

\begin{abstract}
Enabling blockchain technology into IoT can help to achieve a proper distributed consensus based IoT system that overcomes disadvantages of today's centralized infrastructures, such as, among others, high cloud server maintenance costs, weakness for supporting time-critical IoT applications, security and trust issues.
However, meeting requirements posed by IoT in blockchain domain is not an easy endeavour.
\cite{hybridIot} proposes Hybrid-IoT, as a step towards decentralizing IoT with the help of blockchain technology.
Hybrid-IoT consists of multiple PoW sub-blockchains to achieve distributed consensus among IoT devices and an \textit{inter-connector framework}, to execute transactions between sub-blockchains.
In this paper, we take the first step towards designing an inter-connector for multiple blockchains for IoT that is specifically tailored for the Hybrid-IoT architecture. 
We also provide a detailed security discussion, in order to identify threats and we provide discussion on how to cope with  threats.
, 
\end{abstract}


\section{Introduction}\label{sec:introduction}
In 1982, a group of graduate students at  Carnegie Mellon University’s computer science department decided to try to connect their building’s sodas machine to the Internet. Since the the machine was very frequently empty, the students wanted to find a way to remotely check the supply and temperature of sodas, without walking all the way there\footnote{ibm.com/blogs/industries/little-known-story-first-iot-device}. According to many and different sources, this is considered the very first example of an Internet of Things (IoT) device. Today, IoT technology is applied to several environments (e.g. buildings, automotive etc.), with the goal to make them smarter, more connected, profitable, inter-operable and efficient.

Today, most of the IoT platforms are centralized cloud based computing infrastructures. However, their centralized nature presents a number of drawbacks to IoT, such as; high cloud server maintenance costs; weak adoption and support for time-critical IoT applications; security issues (e.g., Single Point of Failure (SPoF)).
On the other hand, decentralization, if achieved, would have the advantage to reduce the amount of data that are transferred to the cloud for processing and analysis, it would be instrumental to improve security and privacy of the managed data \cite{ourPaper}, \cite{ourPaper2}, and it would lead to concerted and autonomous operations.
Decentralized IoT systems have to be able to process high throughput of transactions and scale to many peers in achieving consensus without a trusted central authority. Therefore, IoT decentralization requires frameworks that employ scalable and performant distributed consensus among peers. Lack of such frameworks has been a bottleneck against successful decentralization of many domains including IoT. 
A promising decentralized platform for IoT is blockchain, since it opened a way to overcome distributed consensus bottlenecks in a decentralized setting for large scale applications (cfr. Section \ref{sec:background} for background information on blockchain).

\textbf{Hybrid-IoT.}
Within the blockchain context and approach,  Hybrid-IoT is an ecosystem for IoT decentralization, and it presents a hybrid blockchain architecture that is tailored for IoT, as presented in \cite{hybridIot}. Hybrid-IoT consists of multiple \textit{PoW sub-blockchains} to achieve distributed consensus among IoT devices and an \textit{inter-connector framework}, to execute transactions between sub-blockchains.
Hybrid-IoT design enables IoT devices to operate and collaborate autonomously by by-passing performance and scalability bottlenecks in a decentralized setting.
This opens a way to design autonomous and smart IoT systems that, unlike today's IoT systems, doesn't need any centralized party to perform their tasks.

\textbf{Inter-connector framework.} 
In this work, we present the design of multiple blockchain inter-connector that is specifically tailored for the Hybrid-IoT ecosystem. While the overall Hybrid-IoT idea was introduced in \cite{hybridIot}, that work focused on the design and implementation of sub-blockchains, announcing the design of the inter-connector framework as future work. Proposed inter-connector framework connects many-sub-blockchains as a relay system and its objectives and characteristics are: execute transactions among the sub-blockchains; guarantee transactions  execution, security and accountability; not exercise any direct control over any of the sub-blockchain's consensus mechanism; connect to the sub-blockchains and monitor their execution, and diagnose malfunctioning sub-blockchains.
Proposed design for the multiple blockchain inter-connector consists of three main modules, namely: a \textit{consensus module} (see Section \ref{sec:consensusModule}), a \textit{finality module} (see Section \ref{sec:finalityModule}) 
and a \textit{relay mechanism} (see Section \ref{sec:relayMechanism}). The consensus module consists of an embedded consensus mechanism that executes set of processes (i.e. transactions of sub-blockchains).
The finality module computes the \textit{transaction acceptance times} of inter-chain transactions, which corresponds to the amount of time that needs to pass, while such transactions held by the Hybrid-IoT inter-connector, for inter-chain transactions to be considered as irreversible in their source sub-blockchain.  This is due to the fact that block finality is probabilistic in PoW protocol used by sub-blockchains, and the Hybrid-IoT inter-connector adjusts and uses various mathematical models and algorithms and also takes target security levels between co-operating sub-blockchains, referred to as \textit{collaboration policies}.
The Relay mechanism performs two main tasks: it transfers inter-chain transactions between sub-blockchains to the consensus module for processing; it connects to the to finality module in order to fetch transaction acceptance times for inter-chain transactions;
and, it connects to the consensus module to watch for new transactions and  transfer them to their destined sub-blockchains. 


\textbf{Structure.} The remainder of this paper is organized as follows.
In Section \ref{sec:background}, we provide background information on blockchains.
We review the literature in Section \ref{sec:related}
We detail the design of  Hybrid-IoT inter-connector in Section \ref{sec:architecture}.
In Section \ref{sec:security}, we discuss the security of the proposed design of the multiple blockchain inter-connector.
Section \ref{sec:conclusions} concludes the paper.

\section{Background on Blockchains}\label{sec:background}
Blockchain relies on the concept of a distributed ledger maintained by a peer-to-peer network \cite{blockchainSurvey}. 
Novelty of the blockchain technology lies in its ability to achieve coordination and verification of individual activities carried out by different parties without a centralized authority or trusted third party, that allows decentralization of application execution with concerted and autonomous operations. Blockchain data structure consists of bundled data chunks called blocks, where peers in a blockchain broadcast blocks by exploiting public-key cryptography. Blocks are recorded in the blockchain with exact ordering. A block contains: a set of transactions; a timestamp; a reference to the preceding block that identifies the block's place in the blockchain; an authenticated data structure (e.g., a Merkle tree) to ensure block integrity.\footnote{Block structure varies in different blockchain protocols, here we list the most common elements.} 

Since blockchains are distributed systems, peers of a blockchain network coordinate and co-operate using consensus protocols, such as Proof of Work (PoW), or BFT (Byzantine Fault Tolerant) protocols. PoW consensus mechanisms rely on the condition of doing some computation to prove legitimacy of the performed operation. On the other hand, BFT protocols depend on  state replication between block generators on the processed transactions, in order to be able to operate correctly and achieve consensus in existence of malicious and arbitrarily behaving block generators.
Proof of Stake (PoS) consensus algorithms are popular with the blockchain protocols as well, which can be grouped under the BFT algorithms. 
In PoS, where next generator of the block is selected via combinations of either random selection with their stake (i.e. either their wealth, or age). algorithms

\section{Related Work}\label{sec:related}
In this section, we provide an overview of the prominent proposals, like the inter-chain connector framework, that aims to connect multiple blockchains not only for IoT, please refer to \cite{interledgerApproaches} for detailed discussion.   
Polkadot ecosystem \cite{polkadot} is a multi-chain framework that aims to bring interoperability among multiple blockchains.
It consists of two types of components: \textit{parachains} are blockchains that are tailored according to their tasks and use-cases to computations and operations, and, the \textit{relay chain} is the central chain in the Polkadot ecosystem that performs transactions among parachains.
Polkadot follows a hybrid consensus strategy and splits up block production and finality gadget\footnote{https://wiki.polkadot.network/en/latest/polkadot/learn/consensus/}. 
Moreover, relay chain aims to provide shared security and trust free transactions between parachains.
This requires parachain transactions to be validated by relay chain, which in turn limits flexibility in designing and building parachains.
Another example is Cosmos \cite{cosmos}, which is a blockchain network architecture, similar to the Hybrid-IoT inter-connector and Polkadot that encompasses multiple blockchains, called zones, and provide inter-operability among them through inter-connector called Cosmos Hub.
The Cosmos is based on Tendermint \cite{tendermint} PBFT consensus algorithm.
It separates states of cosmos hub and zones (i.e. cosmos hub and zones as in \cite{cosmos}), and aims to establish a heterogeneous network of Proof of Stake (PoS) zones.
Interledger \cite{interledger} is a protocol for secure transfers across payment systems.
It aims to allow atomic transfer value among any two parties by providing a cryptographically signed receipt from receiver to the sender to not to incur any risks. 
Interledger protocol achieves atomicity of transaction execution by using an adoption Two-Phase commit protocol \cite{twoPhaseCommit} to the financial model as a transaction commit protocol.  
However, Interledger's model requires additional communication between involved parties and notaries that act as transactions managers in the two-phase commit protocol and demands additional resources for executing cryptographic operations. 
Which, considering high volume of transactions to process, is costly for resource constrained IoT devices, unlike the inter-connector design presented in this paper.
\cite{PoWsidechains} presents a decentralized construction for trustless communication between PoW blockchains, as a way to allow passing of different information between blockchains, not only assets.
Similar to the Hybrid-IoT inter-connector, it presents a way to connect PoW blockchains. 
Unlike the multiple blockchain inter-connector protocol presented in this paper, it requires PoW blockchains to have cryptographic primitive NIPoPoWs \cite{nipopows}. 

\section{Blockchain Inter-connector Design}\label{sec:architecture}

Introducing multiple blockchains in an IoT ecosystem requires to generate a collaboration and co-operation infrastructure for transactions processed in different blockchains.

\textbf{Design approach.} 
There are different approaches for designing an inter-connector system. For example, Polkadot consists and progresses a global state of inter-connector and sub-blockchains (i.e. relay chain and parachains as in \cite{polkadot}).
Therefore, the inter-connector validates the states of sub-blockchains, and it is responsible for their correct execution and sub-blockchains benefit from shared security provided by the whole Polkadot ecosystem.
On the other hand, Cosmos \cite{cosmos} separates states of inter-connector and sub-blockchains (i.e. cosmos hub and zones as in \cite{cosmos}), and aims to connect independent blockchains, referred to as \textit{chain sovereignty}. 
Another approach may be leaving blockchains independent from the rest of the system while employing their consensus mechanisms with their own metrics.
In this approach, neither blockchain inter-connector framework nor other blockchains have control over one blockchain's consensus mechanism.
For this approach, blockchain inter-connector must be connected to the blockchains and monitors their execution in order to diagnose blockchains that are malfunctioning (i.e. failing to function properly as defined by the protocol), either because they are under attack (e.g. double spend attack \cite{doubleSpending}), or there is a network congestion (e.g. high stale rates, long block generation intervals etc.).

For Hybrid-IoT ecosystem, we consider the last design approach as the most suitable one, since it is more flexible to be applied in various use cases and thus it is more feasible to adopt to IoT domain.
Therefore, for the rest of this section, we focus on this approach and provide an overview of modules required to realize this design approach and tailor it according to the needs of the Hybrid-IoT ecosystem. 
Designed multiple blockchain inter-connector consist of at least three main elements, namely; \textit{consensus module} (see Section \ref{sec:consensusModule}), \textit{finality module} (see Section \ref{sec:finalityModule})
and \textit{relay mechanism} (see Section \ref{sec:relayMechanism}).

\subsection{Consensus module}\label{sec:consensusModule}
As a distributed system, multiple blockchain inter-connector framework executes set of processes (i.e. executing transactions of blockchains) through a consensus mechanism that is embedded in the \textit{consensus module}.
Multiple blockchain inter-connector framework consists of different machines (i.e. servers) to process and transmit transactions between sub-blockchains.
Consensus module's task is to regulate inter-connector execution and it consists of a consensus mechanism (i.e. implementation of a consensus algorithm).
Consensus mechanism can employ various kinds of consensus algorithms, with various levels of fault tolerance guarantees, such as Crash Fault Tolerant (CFT) algorithms like Raft \cite{raft}, or Byzantine Fault Tolerant algorithms such as \cite{tendermint}, \cite{hyperledger}.
The choice of algorithm should be based on assumption on the network and participants, and their threat model.
Consensus module should be designed as an independent  module that can be plugged to the multiple blockchain inter-connector. This is due to the fact that  operation of the inter-connector framework (e.g. such as how to  connect  to  blockchains,  how  to  transfer  transactions) has to be independent  from the type of consensus mechanism used by the consensus module.

\subsection{Relay Mechanism}\label{sec:relayMechanism}
The relay mechanism have at least three roles: first, listening sub-blockchains to check new blocks, and transfer transactions between sub-blockchains to the consensus module;  second, connecting to the finality module to send network and consensus level statistics and fetching transaction finality times; and third connecting to the consensus module to watch for new recorded transactions to transfer them to their target sub-blockchains.
In order to watch sub-blockchains and the consensus module, the relay mechanism must include set of dedicated parties, referred to as \textit{relay nodes}, to perform transaction transmission among relay mechanism and peers of blockchains. 
Each sub-blockchain has at least one dedicated relay node(s), to transfer it's transactions targeted for another sub-blockchain and receive transactions sent from other sub-blockchains.
Having more than one relay node dedicated to a sub-blockchain increases fault tolerance of the system, since if one node gets faulty, inter-chain transactions can be still performed using another relay node. 
The connection between relay nodes and finality module is twofold: first, relay nodes periodically send network and consensus level statistics to finality module; and second, relay nodes fetch transaction acceptance period for inter-chain transactions from finality time table in finality module.
A dedicated relay node performs its roles by running different processes in parallel, which are: 
connecting to sub-blockchains to fetch inter-chain transactions and monitoring network activities;
sending latest network statistics to finality module and fetching finality time table;  
storing inter-chain transactions in the ledger of the consensus module, that are fetched from their dedicated sub-blockchain;
and, watching ledger of the consensus module for newly written inter-chain transactions sent from other sub-blockchains to it's dedicated sub-blockchain.

\subsection{Finality Module}\label{sec:finalityModule}
When blockchains employ the PoW consensus paradigm, in which block finality is probabilistic, and thus transactions inside the newly mined block can be reverted by generation of conflicting blocks.
This means, at any point of time, another block can be proposed by the network, and if the protocol specific conditions hold (e.g. the new block extends the length of the blockchain and it's timestamp is in the correct timestamp range range like Nakamoto Consensus for Bitcoin\footnote{en.bitcoin.it/wiki/$Block_timestamp$}), the new block can be added and the older block can be pruned from the blockchain. 
However, the possibility of a block being pruned decreases gradually with more blocks connected to it in a chain structure, since they are considered as confirmations. 
As such, in Bitcoin \cite{bitcoinNakamoto}, after six confirmations a block is considered as final (i.e. it is impractical for that block to be removed from the blockchain).
Literature offers interesting approaches that deal with the mathematical modeling of the blockchain protocols with different assumptions on the synchronicity, attacker policy etc., such as \cite{sompolinskyBitcoinSecRevisted, onTheSecurityPoW, analysisOfBlockchainInAsync, bitcoinBackboneProtocolAnalysis}.

The finality module's task is to calculate block finality\footnote{Here finality refers to very high probability, due to the impossibility of finality in Nakamoto Consensus.} time with up to date parameters fetched from the relay mechanism. 
In the inter-connector framework, block finality time translates to the amount of time that should elapse before processing inter-chain transactions inside the block, referred to as \textit{transaction acceptance period}.
Similar to the consensus module, finality module is plugged to the inter-connector, and therefore it is flexible to employ any mathematical model to calculate transaction acceptance period.

The finality module introduces the concept of \textit{collaboration policies}, which are used to model target security level of the inter-chain transaction execution between any two sub-blockchains.
As such, depending on the application scenario, parties in two sub-blockchains may trust each other more than others, and thus they might demand low security level from the finality module which translates to having shorter transaction acceptance periods between those sub-blockchains. 
On the contrary, other sub-blockchains may have a collaboration policy that demands a high level security, which in turn requires more confirmations and longer transaction acceptance period. 
Collaboration policies between any two sub-blockchains are introduced to the finality module when they join to the Hybrid-IoT ecosystem.
The finality module periodically calculates block finality time with up to date input parameters and collaboration policies between sub-blockchains, then, updates and stores them in the \textit{finality time table}.
Relay mechanism uses finality time table to regulate waiting time for transaction execution between sub-blockchains.

Finality module adjusts and uses various mathematical models and algorithms.
As the initial model, we envision to adjust and use the model presented in \cite{sompolinskyBitcoinSecRevisted}, which offers three different robustness notions that correspond to different security guarantees, one of which will be adopted by the finality module to calculate waiting time to execute inter-chain transactions with given collaboration policies.
Therefore, the input parameters of the finality module are: mining power percentage of the devices that are trying to alter the consensus protocol, i.e. attackers\footnote{Here, we presume the worst case, that is all devices that are malfunctioning are collaborating to cheat the system so they are attackers, rather than acting individually, to maximize their success probability.}, and block generation intervals, both fetched from relay nodes; and collaboration policies as target security levels of the inter-chain transaction executions.

\section{Security Discussion}\label{sec:security}
Security vulnerabilities, if not properly addressed and prevented, may allow adversaries to damage correct execution of IoT systems and the proposed multiple blockchain inter-connector. As described in Section \ref{sec:architecture}, the inter-connector takes counter-measures against some of the attacks that might affect sub-blockchains. Such attacks can be grouped as network-level attacks and consensus attacks.
Network-level attacks include: eclipse attacks \cite{eclipseAttack}, network deadlocks, and, Distributed Denial of Service (DDOS) \cite{ddosInTheIoT} attacks.
Consensus-level attacks include: double-spend attacks \cite{doubleSpending}; selfish mining attacks \cite{selfishMining}; and, 50\%+ mining power attacks\footnote{en.bitcoin.it/wiki/$Majority_attack$}.
In the following we discuss the counter-measures will be taken by the inter-connector against those attacks.

Proposed relay nodes have direct connections to all miner IoT devices in the blockchains. This assumption enables relay nodes to have the consensus and network level control by preventing various attack vectors.
As such, first, dedicated relay nodes detect the IoT devices who has not been responding back to its' connection messages, labels that device as a malfunctioning device.
Dedicated relay nodes monitor sub-blockchains to calculate network and consensus level statistics, such as; average block generation interval, network diameter, block and transaction traverse times, and network latency.
Thanks to that, relay nodes have the overall view of the network and can detect malfunctioning devices and network connections that cause network deadlocks and connection problems, and can suggest alternative routes and connections to the devices.
Moreover, dedicated relay nodes can detect eclipse attacks \cite{eclipseAttack} by monitoring IoT devices, and detecting the ones that are not getting any new transactions or blocks that have been published by others, thus targeted by an eclipse attack. 
Dedicated relay nodes can also detect devices that are part of a malicious botnet and performing DDoS attacks, by examining the devices generating extensive amount of network traffic (i.e. through transactions and blocks)  and overall network traffic. 
In order to prevent 50+\% attack (also known as majority attack), relay nodes take a conservative approach and consider all malfunctioning and not-responding devices as attackers, and treats them as a single adversary.
In case of substantial increase in the number of such devices and their percentage of mining power reaches pre-defined thresholds, depending on the application scenario and security metrics ( e.g. when selfish mining attack \cite{selfishMining} is a relevant scenario, than threshold is  set to 33\%), relay nodes stop transferring inter-chain transactions from that sub-blockchain.
Additionally, relay nodes are able to identify selfish mining and associated attacks (e.g. stubborn mining \cite{stubbornMining}), since block and transaction sharing and transmission happens in between members of a closely connected group in the network, dedicated relay nodes extract communication patterns to detect the selfish mining groups.  
Upon detecting malfunctioning and attacker devices, dedicated relay nodes notify all devices in respective sub-blockchain to reject connections and network packets sent by those devices. them, and rejects all transactions sent by those.

We would like to note that, by using a BFT consensus module and by making the consensus module able to check claims of different dedicated relay nodes of a sub-blockchain, we can remove the trust relation in between sub-blockchains to their dedicated relay nodes. This would allow system to be more flexible to the joining and leaving of relay nodes and will have higher level of fault tolerance (i.e. BFT).

\section{Conclusions and Future Work}\label{sec:conclusions}
In this paper, we presented a novel design of the multiple blockchain inter-connector for IoT that is tailored for the Hybrid-IoT ecosystem. It consists of three modules in order to achieve co-operation among multiple blockchains: consensus module, relay mechanism and finality module.
We also provide a detailed security discussion on the proposed design on what are the threats and how they can be addressed.

As future work, first, we plan to define execution algorithm of the relay mechanism and test its performance with performance experiments. 
increase number of experiments we run to test Hybrid-IoT inter-connector's performance.
Second, we plan to prove correctness of the Hybrid-IoT inter-connector and overall Hybrid-IoT system execution with security proofs. 
Finally, we plan to increase resilience of the Hybrid-IoT ecosystem against malicious botnets\footnote{A botnet is a collection of compromised internet computers being controlled remotely by attackers for malicious and illegal purposes \cite{botnetCommunicationPatternsSurvey}, through injection of malicious softwares, malwares, to control them in their command or to steal confidential information.} by adding botnet detection mechanisms such as \cite{autobotcatcher, detectingP2PbotnetsNetworkBehavAnalysisML}).



\bibliographystyle{IEEEtran}
\bibliography{biblio}

\end{document}